\def\rfr#1{eq. (\ref{#1})}

\def\asec{$''$ cy$^{-1}$}

\def\asec{$''$ cy$^{-1}$}

\def\bar{\begin{eqnarray}}
\def\ear{\end{eqnarray}}
\def\bb{\bibitem}
\def\eqi{\begin{equation}}
\def\eqf{\end{equation}}
\def\eqia{\begin{eqnarray}}
\def\eqfa{\end{eqnarray}}
\def\rp#1#2{{#1\over#2}}
\def\ct#1{\cite{#1}}
\def\lb#1{\label{#1}}

\def\oc2{$\mathcal{O}(c^{-2})$}

\def\bds#1{\boldsymbol{#1}}

 \documentclass[11pt]{article}
\usepackage{amsmath,amsthm,amscd,amssymb}
\usepackage{latexsym}
\usepackage{graphicx,epsfig}

\begin{document}

\noindent{\bf \LARGE{Constraining MOND with Solar System dynamics}}
\\
\\
\\
{Lorenzo Iorio}\\
{\it Viale Unit$\grave{a}$ di Italia 68, 70125\\Bari, Italy
\\tel./fax 0039 080 5443144
\\e-mail: lorenzo.iorio@libero.it}

\begin{abstract}
In this letter we investigate the deep Newtonian regime of the MOND paradigm from a purely phenomenological point of view by exploiting the least-square estimated corrections
to the secular rates of the perihelia of the inner and of some of the outer planets of the Solar System by E.V. Pitjeva with the EPM2004 ephemerides. By using $\mu(x)\approx 1-k_0(1/x)^n$ for the interpolating MONDian function, and by assuming that $k_0$, considered body-independent so to avoid violations of the equivalence principle, experiences no spatial variations throughout the Solar System we tightly constrain $n$ with the ratios of the perihelion precessions for different pairs of planets.
We find that the range $1\leq n\leq 2$ is neatly excluded at much more than $3-\sigma$ level. Such a test would greatly benefit from the use of extra-precessions of perihelia independently estimated by other groups as well.
\end{abstract}

PACS: 04.80.-y; 04.80.Cc; 95.10.Ce; 95.10.Eg\\
Keywords: Experimental studies of gravity; Experimental tests of gravitational theories; Celestial mechanics; Orbit determination and improvement

 \section{Introduction}
The MOND scheme (MOdified Newtonian Dynamics) was put forth by Milgrom in Ref.~\cite{Mil83} in order to
phenomenologically explain two basic facts concerning spiral galaxies without resorting to the concept of hidden dark matter: the asymptotic flatness of the rotation curves of spiral galaxies \ct{Rub01} and the Tully-Fisher law which is a well-defined relationship between the rotation velocity in spiral galaxies and their luminosity \ct{Tul77}.

Viewed as a modification of gravity\footnote{It can also be considered as a modification of the inertia of a particle under the action of a generic force $\bds F$.}, MOND predicts that the gravitational acceleration $\bds A_g$ felt by a particle in the field of a distribution of mass is
\eqi\bds A_g=\rp{\bds A_{\rm N}}{\mu\left(\rp{A_g}{A_0}\right)},\eqf where
$\bds A_{\rm N}$ is the Newtonian acceleration, $A_0$ is an acceleration scale which different, independent ensembles of observations set to \ct{San02} $A_0=1.2\times 10^{-10}$ m s$^{-2}$, and $\mu(x)$ is an interpolating function which approximates 1 for $x\gg 1$, i.e. for accelerations larger than $A_0$; for $x\ll 1$ $\mu(x)=x$, so that in such a strongly MONDian regime $A_g\approx \sqrt{A_{\rm N}A_0}$.

Here we wish to investigate the deep Newtonian regime ($x\gg 1$) in view of recent advances in planetary orbit determination occurred for the inner planets of the Solar System.
For a quite general class of interpolating functions, $\mu(x)$
can be cast into the form \ct{Mil83}
\eqi\mu(x)\approx 1-k_0\left(\rp{1}{x}\right)^n,\lb{milgrom}\eqf
which yields a modified gravitational acceleration \ct{Tal88}
\eqi \bds A_g\approx \bds A_{\rm N}\left[1+k_0\left(\rp{A_0}{A_{\rm N}}\right)^n\right].\lb{accmilgrom}\eqf
Note that the most commonly used expressions for $\mu(x)$, i.e. \ct{Mil83}
\eqi\mu(x)=\rp{x}{\sqrt{1+x^2}},\eqf and \ct{Fam05}
\eqi\mu(x)=\rp{x}{1+x},\eqf reduce to \rfr{milgrom} for $k_0=1/2, n=2$ and
$k_0=1, n=1$, respectively, in the appropriate limit. For a recent review of many aspects of MOND as various attempts to theoretically justify it see Ref.~\cite{Bek06} and Ref.~\cite{Bru07}.
\section{Constraints from planetary perihelion precessions}
It can be shown that \rfr{accmilgrom} affects the orbital motion of a test particle in the field of a central mass $M$ with a secular rate of the longitude of the pericenter \ct{Ser06}
\eqi\dot\varpi = -\rp{k_0 n\sqrt{GM}}{r_M^{2n}}a^{2n-\rp{3}{2}}+ \mathcal{O}(e^2)\approx  Q a^z,\lb{peri}\eqf
where $r_M\equiv\sqrt{GM/A_0}$, $a$ and $e$ are the orbit's semimajor axis and eccentricity, respectively, and
\eqi Q\equiv -\rp{k_0 n\sqrt{GM}}{r_M^{2n}},\ z\equiv 2n - \rp{3}{2}.\eqf
The expression of \rfr{peri} can be fruitfully used in conjunction with the corrections to the known  Netonian/Einsteinian secular rates of the perihelia of the inner planets of the Solar System phenomenologically estimated as least-square solve-for parameters in Ref.~\cite{Pit05}
by fitting a planetary data set spanning almost one century with the dynamical force models of the EPM2004 ephemerides \ct{Pit05b}. The same procedure was followed also for some of the outer planets \cite{Pit06}. Since such models fully include Newtonian and Einsteinian gravity, such estimated extra-precessions  account, in principle, for all unmodelled physical effects possibly present in nature like, e.g., MOND. If and when other groups will independently estimate their own extra-precessions of perihelia it will be possible to enlarge and enforce the present test. For a search of other MOND-like effects in the Solar System see Ref.~\cite{BekMag07}, while  the possibility of testing a MONDian violation of the Newton's second law in a terrestrial environment is discussed in  Ref.~\cite{Ign07}. By assuming that MOND does not violate the equivalence principle, i.e. $k_0$ and $n$ are not body-dependent, it is possible to consider for  a generic pair of planets A and B the ratio of their perihelion rates getting\footnote{It is implicitly assumed that $k_0$ does not experience spatial variations, according to the MONDian point of view for which modifications of Newtonian gravity does not depend on distance but on acceleration only \ct{San02}. Of course, our test based on the ratio of the perihelia is valid for the case $k_0\neq 0,n\neq 0$.}
\eqi \rp{\dot\varpi^{(\rm A)}}{\dot\varpi^{(\rm B)}}=\left[\rp{a^{(\rm A)}}{a^{(\rm B)}}\right]^z.\eqf
By defining \eqi \Pi\equiv \rp{\dot\varpi^{(\rm A)}}{\dot\varpi^{(\rm B)}},\eqf
and \eqi\Theta_n\equiv \left[\rp{a^{(\rm A)}}{a^{(\rm B)}}\right]^z,\eqf it is possible to construct
\eqi\Gamma_n\equiv \Pi-\Theta_n;\eqf if, for a given value of $n$, the quantity $|\Gamma_n|$, computed with the extra-rates of perihelia of Ref.~\cite{Pit05} estimated without including any exotic acceleration with respect to standard Newton-Einstein one in the suite of the dynamical force models used to fit the data, turns out to be incompatible with zero within the errors, i.e. if $|\Gamma_n|/\delta\Gamma_n >1$, that value of $n$ must be discarded.  Note that our test makes sense for $k_0\neq 0$, as it is just the case from galactic data.
The uncertainty in $\Gamma_n$ can be conservatively assessed as
\eqi\delta \Gamma_n\leq\delta\Pi + \delta\Theta_n,\eqf with
\eqi\delta\Pi\leq \left|\Pi\right|\left[ \rp{\delta\dot\varpi^{(\rm A)}}{|\dot\varpi^{(\rm A)}|} +
\rp{\delta\dot\varpi^{(\rm B)}}{|\dot\varpi^{(\rm B)}|} \right],\lb{errperi}\eqf
\eqi\delta\Theta_n\leq z\Theta_n\left[ \rp{\delta a^{(\rm A)}}{a^{(\rm A)}} + \rp{\delta a^{(\rm B)}}{a^{(\rm B)}}\right].\eqf
The linear sum of the individual errors in \rfr{errperi} accounts for the existing correlations among the estimated perihelia corrections, which reach a maximum of about 20$\%$ for Mercury and the Earth (Pitjeva, private communication 2005).

\begin{figure}
\begin{center}
\includegraphics[width=13cm,height=11cm]{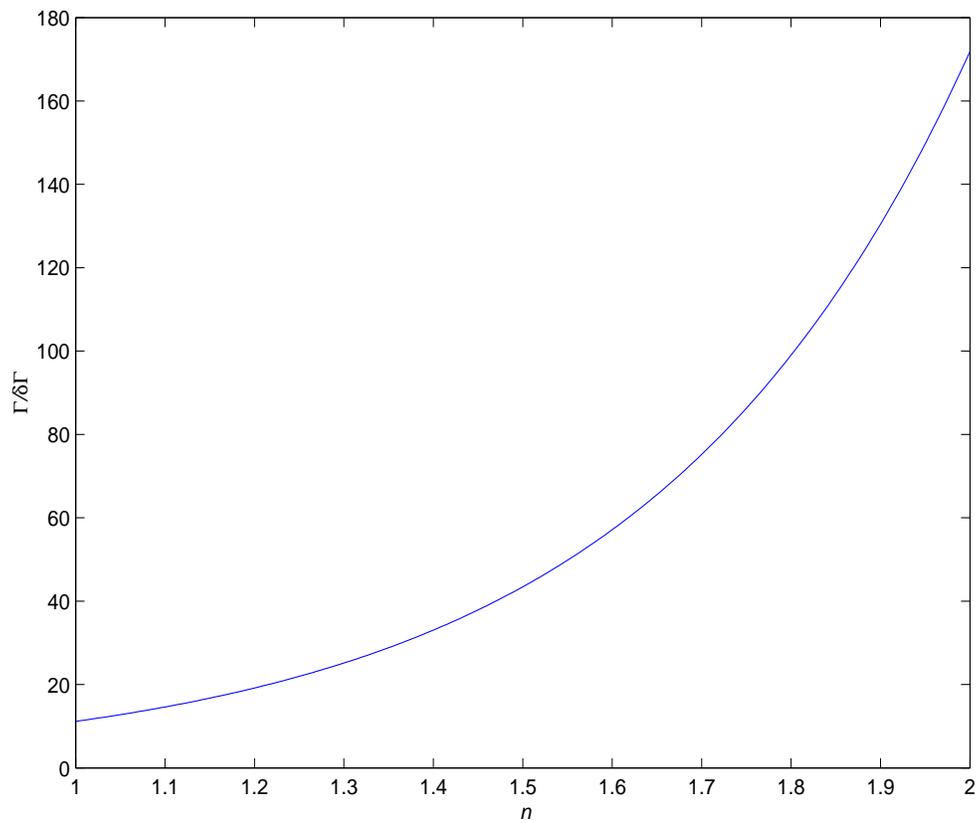}
\end{center}
\caption{\label{figura1} $|\Gamma|/\delta\Gamma$ for Mars and Mercury and $1\leq n\leq 2$. As can be noted, $|\Gamma| \neq 0$ for all values of $n$. }
\end{figure}

\begin{figure}
\begin{center}
\includegraphics[width=13cm,height=11cm]{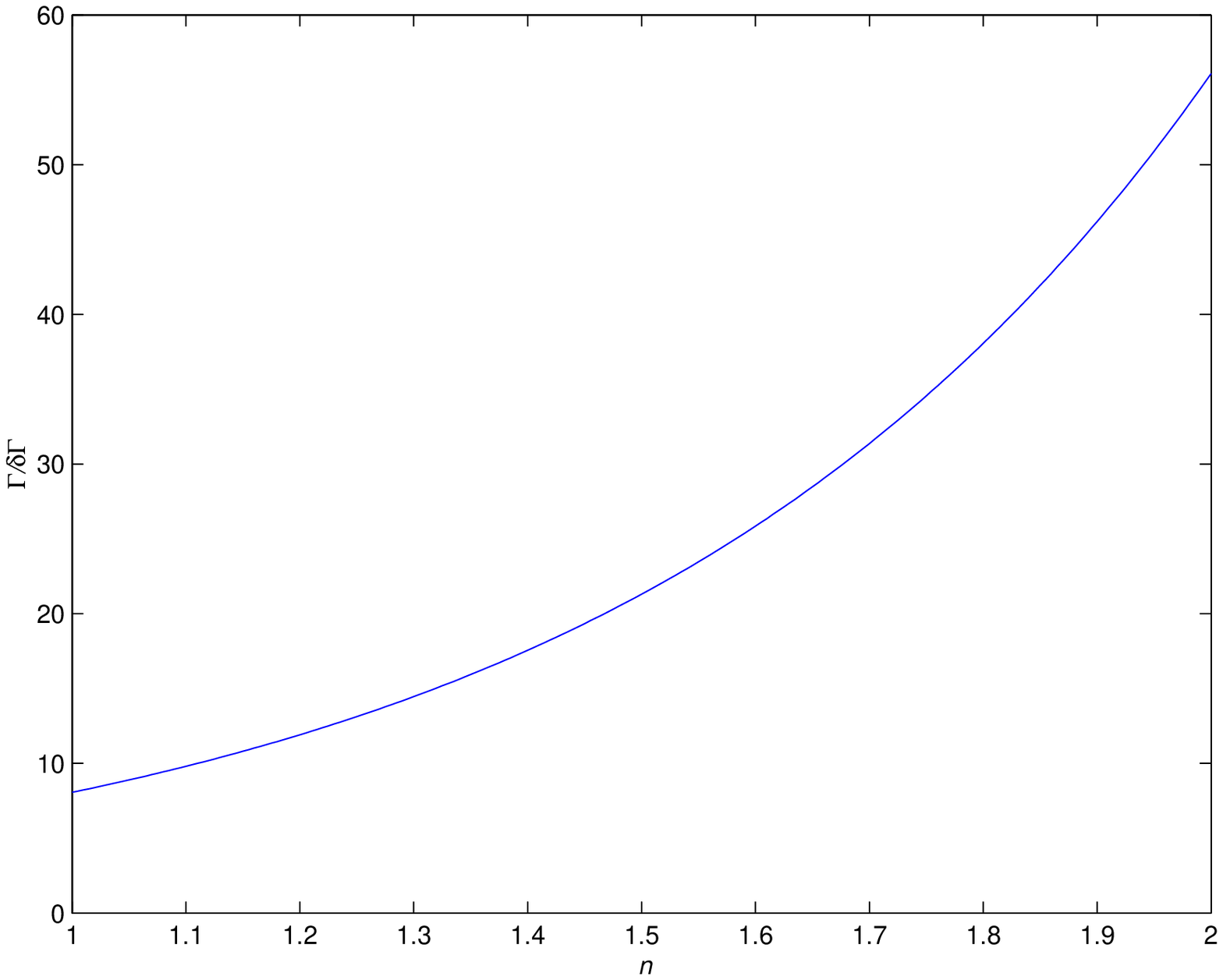}
\end{center}
\caption{\label{figura2} $|\Gamma|/\delta\Gamma$ for the Earth and Mercury and $1\leq n\leq 2$. As can be noted, $|\Gamma| \neq 0$ for all values of $n$.}
\end{figure}

By choosing A=Mars and B=Mercury, from Table \ref{tavola}
\begin{table}[ph]
\caption{Semimajor axes $a$, in AU (1 AU$=1.49597870691\times 10^{11}$ m), and phenomenologically estimated corrections to the Newton-Einstein perihelion secular rates, in arcseconds per century (\asec), of the inner \cite{Pit05} and some of the outer \cite{Pit06} planets. Also the associated errors are quoted: they are in m for $a$ (see Ref.~\cite{Pit05b}) and in \asec\ for $\dot\varpi$ \cite{Pit05, Pit06}. For the semimajor axes they are the formal, statistical ones, while for the perihelia of the inner planets they are realistic in the sense that they
were obtained from comparison of many different
solutions with different sets of parameters and observations (Pitjeva, private communication 2005). The errors quoted here for the perihelia of the outer planets are the formal ones re-scaled by a factor 10 in order to get realistic estimates for them.}
{\begin{tabular}{ @{}ccccc @{} } \hline
 Planet & $a$ (AU) & $\delta a$ (m) & $\dot\varpi$ (\asec)  & $\delta\dot\varpi$ (\asec) \\
\hline
Mercury & 0.38709893 & 0.105 & -0.0036 & 0.0050\\
Earth & 1.00000011 & 0.146 & -0.0002 & 0.0004 \\
Mars & 1.52366231 & 0.657 & 0.0001 & 0.0005\\
Jupiter & 5.20336301 & 639 & 0.0062 & 0.036\\
Saturn & 9.53707032 & 4,222 &  -0.92 & 2.9\\
Uranus & 19.19126393 & 38,484 & 0.57 & 13\\
\hline
\end{tabular}\label{tavola}}

\end{table}
we get Figure \ref{figura1}
in which we plot $|\Gamma|/\delta\Gamma$ for $1\leq n\leq 2$. It turns out that the corrections of order $\mathcal{O}(e^2)$ to \rfr{peri} are negligible in the sense that their inclusion in the calculation does not alter the results for $|\Gamma|/\delta\Gamma$. Moreover,  $\delta\Theta$ is far smaller than $\delta\Pi$ even by re-scaling the formal errors in the semimajor axes by a factor 10 or more. As can be noted, $|\Gamma|$ is always incompatible with zero at much more than $3-\sigma$ level, thus ruling out the interval [1,2] for $n$.   Figure \ref{figura2} refers to A=Earth, B=Mercury: it yields the same conclusions.
It can be shown that the same holds also for A=Mars, B=Venus and A=Earth, B=Venus.
It maybe interesting to note that, although the errors in the perihelion rates quoted in Table \ref{tavola} are not the mere, formal ones, should one decide to re-scale them by a factor 10 our conclusions would remain unchanged, apart from the number of $\sigma$ which would pass from about 180 to 18  (A=Mars, B=Mercury), or from 60 to 6 (A=Earth, B=Mercury) for $n=2$.

Another way to tackle the problem is to construct a $\chi^2$-like quantity $\mathfrak{K}$ defined as follows.
The role of the observables $\mathcal{O}_k$   is played by the ratios of the perihelia precessions $\Pi_k$ for all the $k=1,2,...30$ pairs A/B and B/A of planets, including also the gaseous giant ones whose data are quoted in Table \ref{tavola}. The computed, or predicted, quantities $\mathcal{C}_k$  are the ratios $(a_{\rm A}/a_{\rm B})^{2n-\rp{3}{2}}$  for the corresponding pairs of planets. The errors $\sigma_k$ are $\delta\Pi_k$ because of the negligible impact of the uncertainties of the semimajor axes, while we will assume the number of different pairs constructed, i.e. 30, for the number of degrees of freedom $d$. Thus, we obtain
\eqi \mathfrak{K}=\rp{1}{d}\sum_{k=1}^{30} \rp{(\mathcal{O}_k - \mathcal{C}_k)^2}{\sigma_k^2} \approx 10^3, n=2,\eqf which largely confirms our previous conclusion. It turns out that $\mathfrak{K}\gg 1$ also for $1< n <2$.

\section{Discussion and conclusions}
A major outcome of our analysis is that $n=2$ is neatly ruled out independently of $k_0$ and $A_0$, contrary to what obtained in Ref.~\cite{Ser06} in which only the perihelion of Mars was used by keeping fixed $k_0\approx 1$ (and using the commonly accepted value \ct{San02} $A_0=1.2\times 10^{-10}$ m s$^{-2}$). The results of the analysis presented here seem to suggest that a reconsideration of the matching between the deep Newtonian and MONDian regimes should be looked for. However, caution is in order because the present analysis is based upon the extra-precessions of perihelia estimated by only one team; it would be of great importance to have at disposal  corrections to the perihelia rates determined independently by other groups as well.


\end{document}